\documentclass[showpacs,superscriptaddress,11pt]{revtex4}
\usepackage{graphicx}
\usepackage{amsmath}
\usepackage{amssymb}
\usepackage{latexsym}
\usepackage{array}
\usepackage{amsfonts}
\usepackage{mathrsfs}
\usepackage{verbatim}
\usepackage{multirow}
\usepackage{color}

\begin{document}

\title{Preparation-Attack-Immune Quantum Secure Direct Communication Using Two-Fold Photon Degree of Freedom}

\author{Jihong Min}
\affiliation{Center for Photon Information Processing, and School of Information and Communication, Gwangju Institute of Science and Technology, Gwangju, Republic of Korea}

\author{Jeongho Bang}\email{jbang79@hanyang.ac.kr}
\affiliation{Department of Physics, Hanyang University, Seoul 133-791, Korea}
\affiliation{School of Computational Sciences, Korea Institute for Advanced Study, Seoul 02455, Korea}

\author{B. S. Ham}\email{bham@gist.ac.kr}
\affiliation{Center for Photon Information Processing, and School of Information and Communication, Gwangju Institute of Science and Technology, Gwangju, Republic of Korea}

\received{\today}

\begin{abstract}
Quite recently, enhancing security against device-attack vulnerability has been theoretically challenging but also practically important in quantum cryptographic communication. For dealing with this issue in a general and strict scenario, we design a seemingly-new kind of quantum secure direct communication (QSDC) in a linear-optical regime, which we call ``preparation-attack-immune QSDC.'' We address that in our `naive' analysis, it is quite formidable to extract even a piece of information of the secret message, and any malicious eavesdropping attempts will be unsuccessful. The most remarkable feature is that even in the case where a powerful eavesdropper can peep at all preparation device settings, our protocol still keeps a high level of security without invoking any additional resources and physical properties. This novel advantage that we call `preparation-attack immunity' comes from the simultaneous use of the two degrees of freedom involved in a single-photon (polarization and spatial modes), which enables one to faithfully deal with the single-photon quantum superposition nature. Our idea can be generalized to other single-photon based protocols.
\end{abstract}

%\pacs{LaLa}

\maketitle

\newcommand{\bra}[1]{\left<#1\right|}
\newcommand{\ket}[1]{\left|#1\right>}
\newcommand{\abs}[1]{\left|#1\right|}
\newcommand{\expt}[1]{\left<#1\right>}
\newcommand{\braket}[2]{\left<{#1}|{#2}\right>}
\newcommand{\commt}[2]{\left[{#1},{#2}\right]}
\newcommand{\tr}[1]{\mbox{Tr}{#1}}

For several years a great number of quantum communication protocols has been proposed and studied. However in most cases the security is guaranteed on the condition that the legitimate communication parties (say, Alice and Bob) can trust the devices to initially prepare the valid information carriers, i.e., flying qubits \cite{Hockney03,Qi15,Cao16}. For example, in a conventional quantum communication scheme, the states of the mutually unbiased bases (e.g., eigenstates of $\hat{\sigma}_z$ or $\hat{\sigma}_x$) are commonly used as the information carriers. Then, it should be assumed that Alice and Bob prepare them by using the {\em trustworthy} devices free from connection with any other malicious eavesdropper. However, such an assumption regarding device-trustworthiness or device-reliability cannot always be made due to the room for possibility of hacking those devices or softwares being compromised beforehand. The now accepted way to approach this problem is to exploit the Bell inequality violation with the entanglement distributed between the communicating mates, e.g., in the device-independent scenarios (see ref.~\cite{Acin07}, and refs.~\cite{Tang16, Yin16} for recent results). Still, we are far from the faithful design and realization of such a task, as probably it is quite difficult to generate and distribute the entanglement between far-away places. Thus we feel that it would be fair to find a moderate-step solution taking into consideration the feasibility of current technology~\cite{Lucamarini05,Hu15}.

To deal with such an issue and make our idea general in a strict scenario, we consider here an important branch of the quantum communication scheme, called quantum secure direct communication (QSDC) \cite{Kim02,Deng03,Deng04,Wang05,Wang06,Jie07,Han08,Chang13,Mi15}, which allows two parties Alice and Bob to communicate directly in secret without performing any additional tasks, e.g., a pre-distribution of the secret-keys. The QSDC scheme has received interest as it appears to be efficient \cite{Wang05,Song12,Farouk15} and may result in some advantages with any distributed quantum tasks \cite{Tseng12,Bang15} or quantum network \cite{Hong12}. Here we note that QSDC is substantially different from the quantum key distribution (QKD) in which the two remote users create the private key, and then communicate their messages via classical channel \cite{Gisin02}. The security requirements for QSDC are also stricter than those for QKD (or other quantum cryptographic schemes), because none of the useful information of the secret message should leak out to an eavesdropper; once the eavesdropper has already extracted any information of the secrete message, the QSDC protocol may be of no use even if the eavesdropper can be identified. Thus, we will present an explicit realization of our idea based on using this QSDC scheme, preserving the essential content and general applicability.

Keeping the above-described in mind, we design and propose a ``preparation-attack-immune QSDC'' in a linear-optical framework. The proposed QSDC scheme runs with the photons moving along the pathways `{Alice $\to$ Bob}' and `{Bob $\to$ Alice}.' More specifically, a single-photon generated on Bob's side is thrown to Alice, and it can be reflected---without being assisted by the quantum memory---to Bob with a bit of Alice's message and/or the potential warning on the existence of any malicious eavesdropper, say Eve. The photon could also be caught in one of Alice's detectors, in which this case, Alice and Bob get another chance to detect Eve. On the basis of the naive analysis, we address that it is {\em probabilistically unsuccessful and impractical} to eavesdrop on even one bit of the message due to the complications of the protocol. In particular, the most remarkable feature is that the security of our protocol is not seriously damaged even considering a powerful Eve who can look at Bob's whole preparation settings, whereas other single-photon based QSDC schemes (and most existing quantum cryptographic protocols) may become insecure. This new kind of advantage, called preparation-attack immunity, originates from the simultaneous use of the two degrees of freedom of the single-photon that are the polarization and path modes. We indicate that our idea can be generalized to other single-photon based QKD protocols.

%--------------------------------------------------------------------------------------------
\section*{RESULTS}
%--------------------------------------------------------------------------------------------

First, we briefly describe the scenario of our QSDC scheme. Alice intends to securely deliver a secret message to Bob. The scenario and method follows those of the conventional QSDC schemes designed based on using the single-particle quantum carriers. Bob generates a quantum particle in a state $\in \{\ket{0}, \ket{1}, \ket{+}, \ket{-}\}$ and sends it to Alice through a quantum channel ${\cal C}$. Here, $\ket{\pm} = (\ket{0}\pm\ket{1})/\sqrt{2}$. Then, Alice encodes a bit of her message by applying a unitary $\in \{ \hat{\openone}, \hat{\sigma}_y \}$ (we call ``message operation'' hereafter) to the incoming particle and reflects it to Bob through the same channel ${\cal C}$. Whether or not the state of the returning particle is flipped corresponds to the bit of the message from Alice: `$0$' when Alice acts $\hat{\openone}$, and otherwise, `$1$' when Alice does $\hat{\sigma}_y$. In this work, we have shown that an eavesdropper, Eve, cannot decode the message bits even in the case where all information of Bob's preparation and encoding is leaked, in contrast to the previous schemes \cite{Zhu06,Li06,Wang06}---in this sense, we call our scheme ``preparation-attack-immune QSDC.''

\begin{figure}
\centering
\includegraphics[width=0.85\textwidth]{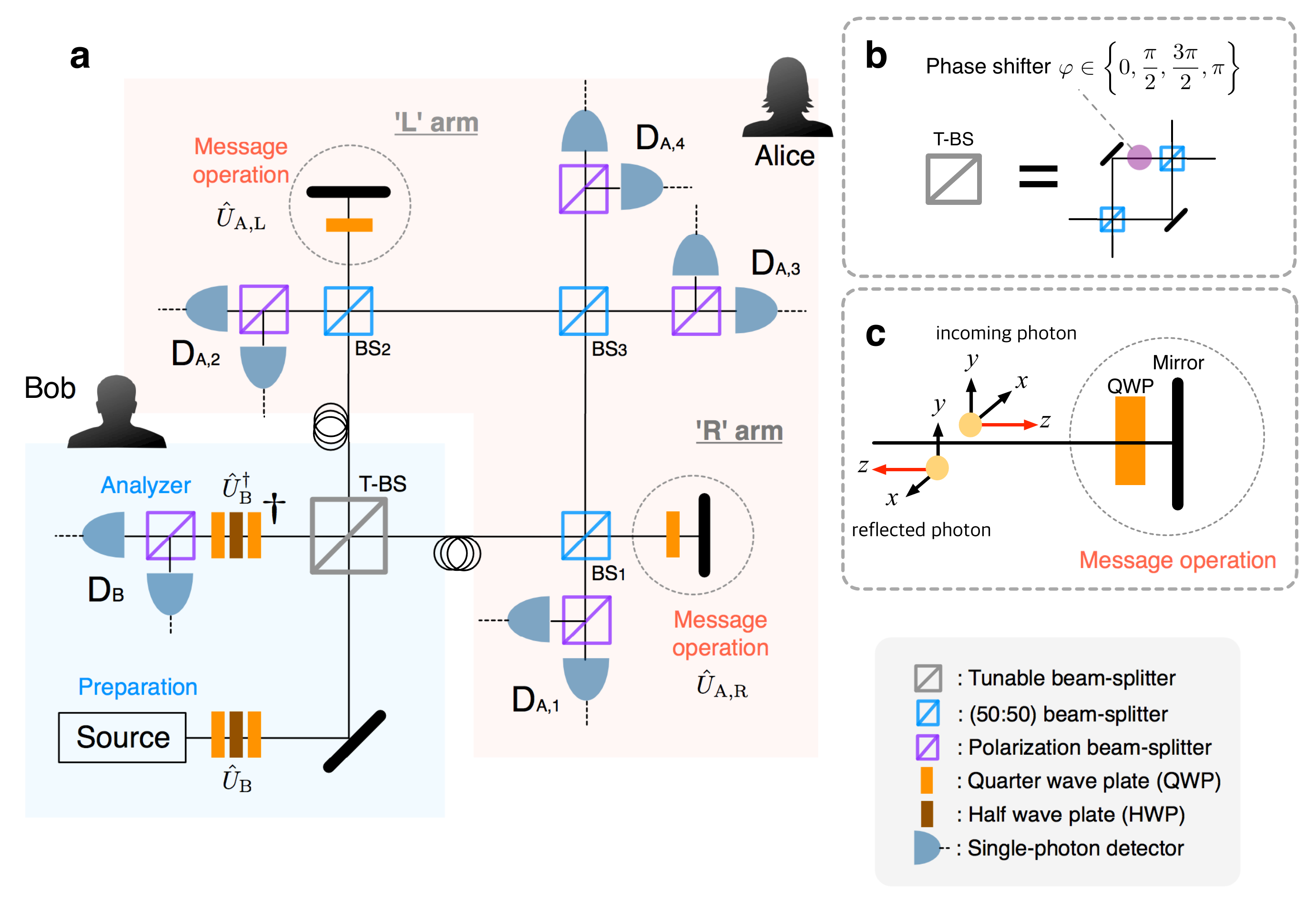}
\caption{\label{fig:main_scheme} (a) The linear-optical setting of our preparation-attack-immune QSDC. Bob owns two modules of QWP-HWP-QWP: One is used to prepare a single-photon in a specific polarized state $\in \{ \ket{H}, \ket{V}, \ket{D}, \ket{A} \}$, and the other is responsible for the analysis of the returning photon. (b) Bob also has a `tunable' beam-splitter (T-BS) to control the photon pathways: R (right), L (left), or their superposition. (c) Alice equips each arm (R or L) with a QWP and a mirror to implement the message operations $\in \{ \hat{\openone}, \hat{\sigma}_y \}$ [as in Eq.~(\ref{eq:message_op})]. The polarization detection modules $\text{D}_{\text{A},n}$ ($n=1,2,3,4$) are placed in Alice's side to find any eavesdropper (if any). Alice and Bob also employ a (classical) public channel ${\cal C}_\text{public}$. Bob prepares a photon and throws it to Alice. The photon can be captured in one of Alice's detecting modules $\text{D}_{\text{A},n}$ ($n=1,2,3,4$) or be reflected to Bob taking a bit of Alice's message (see the main text for detailed analyses).}
\end{figure}  

\subsection{Linear-optical setting.} %-------------------------

We then describe the linear-optical setting of our preparation-attack-immune QSDC protocol. Firstly, let us describe the setup which consists of a single-photon source/detector, beam-splitter (BS), polarization beam-splitter (PBS), and quarter/half wave-plates (QWP/HWP) [see Fig.~\ref{fig:main_scheme}(a)]. Bob has a single-photon preparation module, and a `tunable' beam-splitter (T-BS) which is nothing but a Mach-Zehnder interferometer having a tunable phase $\varphi \in \left\{0, \frac{\pi}{2}, \frac{3\pi}{2}, \pi \right\}$ [see Fig.~\ref{fig:main_scheme}(b)] \cite{Cerf98}. A polarization analyzer $\text{D}_\text{B}$ is also equipped on Bob's side to decode the bits of Alice's message or to check for the existence of Eve's eavesdropping. On the other hand, Alice has four H/V polarization detection modules, denoted by $\text{D}_{\text{A},n}$ ($n=1,2,3,4$). Alice also employs the quarter wave-plate $\text{QWP}({\vartheta})$ and a mirror to implement a message operation $\hat{\openone}$ or $\hat{\sigma}_y$, where $\vartheta$ are the real controllable parameters (i.e., rotation angles) of QWP. Here we assume that $\vartheta$ can precisely be controlled to be either $\frac{\pi}{2}$ or $\frac{\pi}{4}$ for the implementations of $\hat{\openone}$ and $\hat{\sigma}_y$, respectively (as will be described later). This module is placed on each pathway (hereafter called ``arm''). We represent the message operations as $\hat{U}_{\text{A},x}$, where $x=\text{R}, \text{L}$ (R: right, L: left) denotes the direction in which the arm is placed. Alice and Bob are also allowed to use a (classical) public channel ${\cal C}_\text{public}$.

\subsection{Protocol.} %-------------------------

Our protocol runs as follows: First, Bob declares the commencement to Alice through ${\cal C}_\text{public}$, and Alice encodes a bit of her message by setting $\vartheta \in \left\{\frac{\pi}{2}, \frac{\pi}{4}\right\}$ of QWP in either right or left arm (say, R-arm or L-arm). In other words, Alice sets her QWPs to be either 
\begin{eqnarray}
&& \left( \hat{U}_\text{A,R} \in \left\{\hat{\openone}, \hat{\sigma}_y \right\}~\&~\hat{U}_\text{A,L}=\hat{\openone} \right) \nonumber \\
&& ~\text{or}~ \left( \hat{U}_\text{A,R}=\hat{\openone}~\&~\hat{U}_\text{A,L} \in \left\{\hat{\openone}, \hat{\sigma}_y \right\}\right).
\label{eq:message_op}
\end{eqnarray}
To provide a more detailed explanation, the polarization of the photon experiencing the QWP (twice) with the mirror in Alice's module evolves as $\ket{\cal P} \xrightarrow{\text{QWP}(\vartheta) \text{-} \hat{\sigma}_z \text{-} \text{QWP}(-\vartheta)} \ket{\cal P'}$, where ${\cal P}, {\cal P'} \in \{H, V, D, A\}$ and the Pauli matrix $\hat{\sigma}_z = \left( \begin{smallmatrix} 1 & 0 \\ 0 & -1\end{smallmatrix} \right)$ corresponds to the operation of the mirror that changes the ordinary axis of the polarization [see Fig.~\ref{fig:main_scheme}(c)]. Here, the operation of QWP($\vartheta$) is given as
\begin{eqnarray}
\hat{U}_\text{QWP}(\vartheta)=\frac{1}{\sqrt{2}}
\begin{pmatrix}
1-i \cos{2\vartheta} & -i \sin{2\vartheta} \\
-i \sin{2\vartheta} & 1+i \cos{2\vartheta}
\end{pmatrix}.
\end{eqnarray}
Then, the general form of the message operation can be represented as
\begin{eqnarray}
\hat{U}_\text{A, (R or L)} &=& \hat{U}_\text{QWP}(-\vartheta) \hat{\sigma}_z \hat{U}_\text{QWP}(\vartheta) \nonumber \\
&=& -i
\begin{pmatrix}
\cos{2\vartheta} & \sin{2\vartheta} \\
-\sin{2\vartheta} & \cos{2\vartheta}
\end{pmatrix}.
\label{eq:message_U}
\end{eqnarray}
Therefore we can implement the message operation $\hat{U}_\text{A, (R or L)}$ as in Eq.~(\ref{eq:message_op}) (ignoring the global phase factor), such that 
\begin{eqnarray}
\hat{U}_\text{A, (R or L)} =
\left\{
\begin{array}{ll}
\hat{\openone} & \text{if Alice sets}~\vartheta=\frac{\pi}{2}, \\
\hat{\sigma}_y  & \text{if Alice sets}~\vartheta=\frac{\pi}{4}. 
\end{array}
\right.
\label{eq:option_Um}
\end{eqnarray}
After finishing their ready-sign, Bob generates a single photon $\in \left\{ \ket{H}, \ket{V}, \ket{D}, \ket{A} \right\}$ (at random), where $\ket{D}=\left( \ket{H} + \ket{V} \right)/\sqrt{2}$ and $\ket{A}=\left( \ket{H} - \ket{V} \right)/\sqrt{2}$. Such an arbitrary preparation can generally be done by a combination of QWP-HWP-QWP (However, without loss of generality, only two wave-plates, QWP-HWP, would suffice in our case) \cite{Damask04}. Then, Bob sends the prepared photon to Alice. Here Bob chooses one of the four possible ways to transmit the photon, each of which is determined by the setting $\varphi \in \left\{0, \frac{\pi}{2}, \frac{3\pi}{2}, \pi \right\}$ of T-BS. More specifically, if Bob sets $\varphi=0$ or $\varphi=\pi$, then the photon passing through T-BS is either to be transmitted toward R-arm or to be reflected toward L-arm {\em deterministically}. On the other hand, if $\varphi=\frac{\pi}{2}$ or $\varphi=\frac{3\pi}{2}$, the photon can travel in both R-arm and L-arm, simultaneously; in other words, here the two paths are {\em superposed} in these cases. In fact, such a ``superposition path'' plays an important role in our protocol. 
%Here we significantly note that the choice of $\varphi$ is also made at random, {\em which is not opened even for Bob}.

\subsection{Photon traveling paths.} %-------------------------

From now on we describe the process and all occurring events in the two kinds of photon travel paths.

($i$) {\em Deterministic path $(\varphi=0$ and $\varphi=\pi)$.} -- We first consider that the prepared photon goes along the deterministic paths toward either R-arm (i.e., $\varphi=0$) or L-arm (i.e., $\varphi=\pi$). Let us first consider that the traveling photon passes through the beam-splitter $\text{BS}_1$ or $\text{BS}_2$ for each $\varphi$-setting. In such settings, Bob can receive back the reflected photon. Note that if $\varphi = 0$ or $\varphi = \pi$, the reflected photon must come into the analyzer due to the trait of T-BS. In this case, Bob publicly announces only the fact that the photon returns back, and Alice announces which arm (R or L) the valid message operation is carried out in. Here, if the initial path of the photon is matched to the arm announced by Alice, Bob can decode the bit of Alice's message by analyzing whether the initially prepared polarization is flipped or not. Otherwise, in the not-matched case, the measured polarization in Bob's analyzer is to be used to check if there exists Eve's eavesdropping in the way of Bob $\to$ Alice or Alice $\to$ Bob, because the polarization should be unchanged in this case. Another possibility when $\varphi=0$ or $\varphi=\pi$ is that the reflected photon could also be caught by $\text{D}_{\text{A},1}$ placed in R-arm (when $\varphi=0$) or by $\text{D}_{\text{A},2}$ placed in L-arm (when $\varphi=\pi$). Then, Alice can presume the original polarization of the incoming photon (before the message operation is carried out) using $\hat{U}_{\text{A,R}}^\dagger$ or $\hat{U}_{\text{A,L}}^\dagger$. Alice announces the detector number $n$ ($1$ or $2$) and the presumed original polarization through ${\cal C}_\text{public}$. Then, Bob can check if Eve disturbed the photon moving in the way of Bob $\to$ Alice. If the detector announced by Alice is placed on an invalid arm for the $\varphi$-setting, it is indicative of Eve's existence. Here, if the photon is initially prepared as $\ket{H}$ or $\ket{V}$, Bob can do an additional check by comparing the initially prepared polarizations with the later ones identified by $\text{D}_{\text{A},1}$ or $\text{D}_{\text{A},2}$. Other possibilities are that the photon is detected in $\text{D}_{\text{A},3}$ or $\text{D}_{\text{A},4}$. In such cases, Bob can check if the initial state has been altered when his original preparation was $\ket{H}$ or $\ket{V}$.

($ii$) {\em Superposition path $(\varphi=\frac{\pi}{2}$ and $\varphi=\frac{3\pi}{2})$.} -- We investigate the case where the prepared photon follows the superposition path. Firstly, let us consider that the photon can be measured in $\text{D}_{\text{A},3}$ or $\text{D}_{\text{A},4}$. In these two cases, Alice announces the detector number $n$ and the measured polarization, and then Bob checks the possible existence of Eve who could be in the way of Bob $\to$ Alice by analyzing the announced number $n$. Here, note that the photon has to have appeared in $\text{D}_{\text{A},3}$ when $\varphi=\frac{\pi}{2}$ and in $\text{D}_{\text{A},4}$ when $\varphi=\frac{3\pi}{2}$. If the initial state is in $\ket{H}$ or $\ket{V}$, Bob does the additional check of whether the polarization is altered. We can also consider that the photon touches neither $\text{D}_{\text{A},3}$ nor $\text{D}_{\text{A},4}$, but is detected in $\text{D}_{\text{A},1}$ or $\text{D}_{\text{A},2}$. In that case, Alice announces her detection (the number $n$ and the presumed Bob's original polarization) according to the rule described in ($i$). Then Bob has a chance to check the alteration of the polarization, if the photon is initially prepared as $\ket{H}$ or $\ket{V}$. The last possibility is that the photon returns back after passing through all of Alice's detectors. In this case, the photon would just be discarded.

\begin{figure}
\centering
\includegraphics[width=0.95\textwidth]{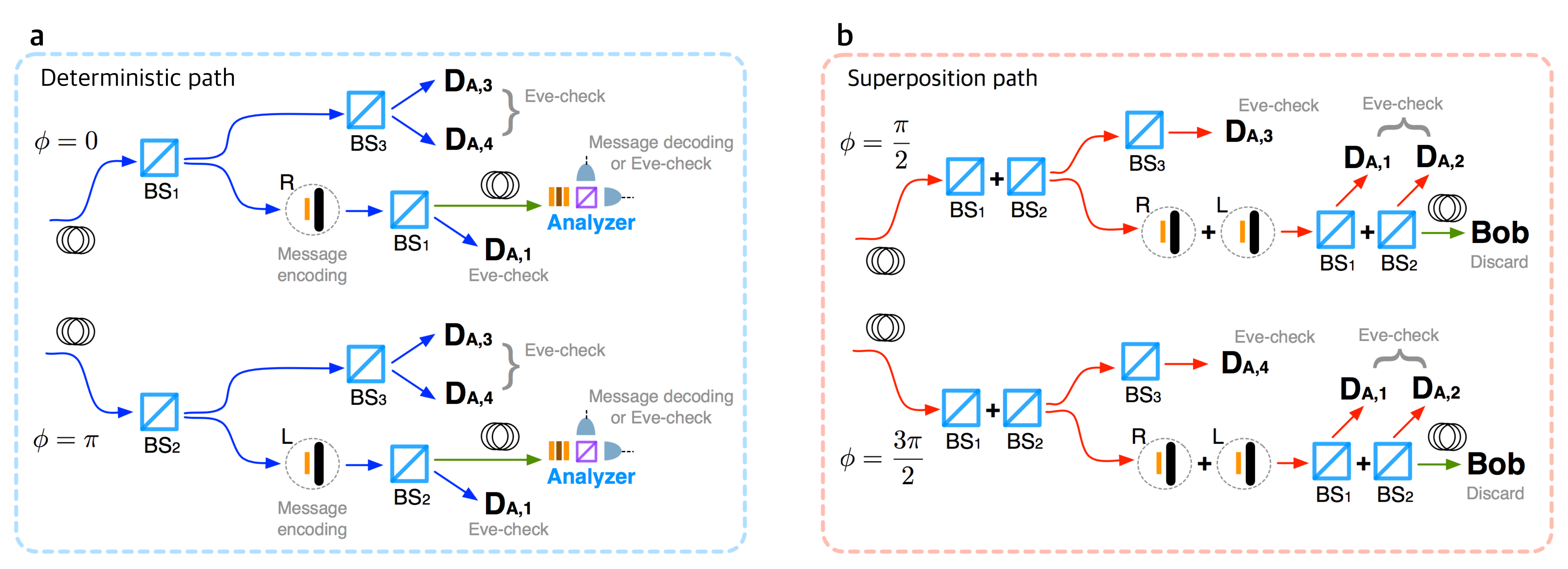}
\caption{\label{fig:tree} All possible events of a trial of our scheme is presented with respect to (a) deterministic and (b) superposition paths. Depending on the chosen $\varphi \in \left\{0, \frac{\pi}{2}, \frac{3\pi}{2}, \pi \right\}$, the photon is supposed to have appeared in one of Alice's detection modules $\text{D}_{\text{A},n}$ ($n=1,2,3,4$), or reflected toward Bob again following the pathways out of the Alice's detection modules. The photon could also be discarded without any contribution to the detection process.}
\end{figure}

For the single trial described as in the above ($i$) or ($ii$), we present a schematic diagram of all possible events in Fig.~\ref{fig:tree}. Alice and Bob repeat the trials until the full message is transmitted with no indication of external alterations. 

\subsection{Event probabilities.} %-------------------------

First we evaluate the event probabilities (by observing Fig.~\ref{fig:tree}) and make the following analyses not taking into account the channel errors and imperfections of the implementing devices. The photon delivers a bit of Alice's message with probability $P_\text{message}=\frac{1}{16}$, or is used to detect Eve with probability $P_\text{Eve-check}=\frac{5}{8}$. The photon could be wasted with probability $P_\text{discard}=\frac{5}{16}$, giving no contribution. Here, it is worth noting that it is possible to reduce the overall time to complete the protocol by increasing the probability $P_\text{message}$ of the message encoding. It can actually be done by favorably dialing the reflectance $r$ (or the transmittance $t=1-r$) of the beam-splitters $\text{BS}_{1}$ and $\text{BS}_{2}$, instead of using a conventional $50$:$50$ (i.e., $r=t=\frac{1}{2}$). To show this more explicitly, we can use the mappings of the photon creation operators in each modes with respect to $\varphi$:
\begin{widetext}
\begin{eqnarray}
\hat{a}_{\text{T-BS,R}}^\dagger &\to& \left(1-r\right) \times \left(\text{\small - reflected toward Bob \& measured -}\right) \nonumber \\
&& + \sqrt{r \left( 1-r \right)} \hat{a}_{\text{BS}_1,\text{D}_\text{A,1}}^\dagger + \sqrt{\frac{r}{2}} \left( \hat{a}_{\text{BS}_3,\text{D}_\text{A,3}}^\dagger - \hat{a}_{\text{BS}_3,\text{D}_\text{A,4}}^\dagger \right)~\left(\text{for $\varphi=0$}\right), \nonumber \\
\frac{\left(\hat{a}_{\text{T-BS,R}}^\dagger + \hat{a}_{\text{T-BS,L}}^\dagger\right)}{\sqrt{2}} &\to& \left(1-r\right) \times \left(\text{\small - reflected toward Bob \& discarded -}\right) \nonumber \\
&& + \sqrt{\frac{r\left(1-r\right)}{2}} \left( \hat{a}_{\text{BS}_1,\text{D}_\text{A,1}}^\dagger + \hat{a}_{\text{BS}_2,\text{D}_\text{A,2}}^\dagger \right) - \sqrt{r} \hat{a}_{\text{BS}_3,\text{D}_\text{A,4}}^\dagger~\left(\text{for $\varphi=\frac{\pi}{2}$}\right), \nonumber \\
\frac{\left(\hat{a}_{\text{T-BS,R}}^\dagger - \hat{a}_{\text{T-BS,L}}^\dagger\right)}{\sqrt{2}} &\to& \left(1-r\right) \times \left(\text{\small - reflected toward Bob \& discarded -}\right) \nonumber \\
&& - \sqrt{\frac{r\left(1-r\right)}{2}} \left( \hat{a}_{\text{BS}_1,\text{D}_\text{A,1}}^\dagger - \hat{a}_{\text{BS}_2,\text{D}_\text{A,2}}^\dagger \right) - \sqrt{r} \hat{a}_{\text{BS}_3,\text{D}_\text{A,3}}^\dagger~\left(\text{for $\varphi=\frac{3}{2}\pi$}\right), \nonumber \\
\hat{a}_{\text{T-BS,L}}^\dagger &\to& -\left(1-r\right) \times \left(\text{\small - reflected toward Bob \& measured -}\right) \nonumber \\
&& + \sqrt{r \left( 1-r \right)} \hat{a}_{\text{BS}_2,\text{D}_\text{A,2}}^\dagger - \sqrt{\frac{r}{2}} \left( \hat{a}_{\text{BS}_3,\text{D}_\text{A,3}}^\dagger + \hat{a}_{\text{BS}_3,\text{D}_\text{A,4}}^\dagger \right)~\left(\text{for $\varphi=\pi$}\right). 
\label{eq:BS_mapping}
\end{eqnarray}
\end{widetext}
where $\hat{a}_{\text{T-BS,R}}^\dagger$ and $\hat{a}_{\text{T-BS,L}}^\dagger$ create a photon as the output of T-BS traveling along R-arm and L-arm; $\hat{a}_{\text{BS}_j,\text{D}_{\text{A},k}}^\dagger$ denotes the operator to create an output photon from $\text{BS}_j$ ($j=1,2,3$) toward $\text{D}_{\text{A},k}$ ($k=1,2,3,4$). From the mappings in Eq.~(\ref{eq:BS_mapping}), we can give the general form of the event probabilities (see Sec.~S1 of the Supplementary Information for detailed derivations),
\begin{eqnarray}
P_\text{message} &=& \frac{\left(1-r\right)^2}{4}, \nonumber \\
P_\text{Eve-check} &=& \frac{1 + 4r - 2r^2}{4}, \nonumber \\
P_\text{discard} &=& \frac{2 - 2r + r^2}{4},
\label{eq:prob_events}
\end{eqnarray}
in which we assumed that the probabilities of choosing $\varphi \in \{ 0, \frac{\pi}{2}, \frac{3\pi}{2}, \pi \}$ are all $\frac{1}{4}$. Then, we can make the following observation: First, if we do not use $\text{BS}_1$ and $\text{BS}_2$ (or equivalently, $r=0$), Alice and Bob have $P_\text{message} = \frac{1}{4}$, $P_\text{Eve-check}=\frac{1}{4}$, and $P_\text{discard}=\frac{1}{2}$. However, as Bob does not need to use the superposition path in this case, the actual event probabilities are $P_\text{message} = \frac{1}{2}$, $P_\text{Eve-check}=\frac{1}{2}$, and $P_\text{discard}=0$. Such a setting, of course, runs effectively and is efficient in the sense that there are no discarded photons. In fact, this setting can be regarded as a combination of two BB84 protocols using T-BS. However, note that it may become insecure against the eavesdropping of Eve who can peep at Bob's polarization encoding. This means that the critical advantage, i.e., the preparation-attack immunity (will be described below), cannot be achieved just by adding up the effective arms. Next, we can also consider a case where Alice replaces $\text{BS}_1$ and $\text{BS}_2$ to the reflecting mirrors (or equivalently, $r=1$). In this case, Alice cannot encode her bit-message on the photon with $P_\text{message}=0$. Bob then does not need to use the deterministic paths, and the photon is not wasted with $P_\text{Eve-check}=1$. Such a setting can be used just to check the security before the full-scale message communication.
%For 50:50 beam-splitters (which happens when $r=\frac{1}{2}$), Alice and Bob have $P_\text{message} = \frac{1}{16}$, $P_\text{Eve-check}=\frac{5}{8}$, and $P_\text{discard}=\frac{5}{16}$. 

\subsection{Impracticality of eavesdropping.} %-------------------------

For the analysis of security, we assume here that there exists a malicious Eve who attempts to eavesdrop on the secret message. We discuss that an attempt by Eve will be unsuccessful due to the following complications:
\begin{itemize}
\item[({\bf C.1})] Firstly, it is assumed that the critical information, such as the initial photon polarization, the setting of $\varphi$, and the position of valid message operation (to be either R or L), are not opened. 
\item[({\bf C.2})] In case Bob prepares $\ket{H}$ or $\ket{V}$, Eve who alters the polarization of the photon moving `{Bob $\to$ Alice}' can be found in $\text{D}_{\text{A},n}$ ($n=1,2,3,4$). Eve who does it in the way of `{Alice $\to$ Bob}' can also be detected by Bob's analyzer $\text{D}_\text{B}$. 
.
\item[({\bf C.3})] Alice and Bob can sense Eve when the photon has appeared in an invalid arm for the initial $\varphi$-settings. In particular, we note that the use of the superposition paths allows to detect even a more powerful Eve who can peep Bob's initial preparation.
\end{itemize}
Based on the above-described complications ({\bf C.1})-({\bf C.3}), it is believed that Eve's eavesdropping is probabilistically unsuccessful and impractical.

Nevertheless, we should always consider the existence of a potential Eve who has a reasonable (powerful) strategy to carry out the eavesdropping within the rules of quantum theory. Thus, we now perform a `naive' theoretical analysis just to provide a convincing proof of the validity and security of our QSDC protocol. For the purpose, our analysis is focused on the Eve's eavesdropping probability that allows to estimate the amount of information Eve has gained. First, let us assume that Eve can adopt the primitive strategy of disturbing the moving photon (e.g., an intercepting-and-resending attack) on the way of `{Bob $\to$ Alice}' and `{Alice $\to$ Bob}.' In such a strategy, Eve needs to see the polarization of the photon moving both forward (i.e., `{Bob $\to$ Alice}') and backward (i.e., `{Alice $\to$  Bob}') ways. Thus, Eve would place her own detecting modules $\text{D}_\text{Eve,R}$ and $\text{D}_\text{Eve,L}$ in between T-BS and Alice. Here, $\text{D}_\text{Eve,R}$ and $\text{D}_\text{Eve,L}$ are set to be either H/V or D/A measurement. Then, Eve is supposed to find a photon in either $\text{D}_\text{Eve,R}$ or $\text{D}_\text{Eve,L}$. After the analysis of the measured photon, Eve should resend the photon (possibily randomly polarized) without attracting the attention of Alice and/or Bob. Here it is assumed that Eve already understands the protocol. Then, by considering all possible events described in Fig.~\ref{fig:tree}, we can evaluate the probability that Eve successfully eavesdrops on one bit of the message with respect to the reflectivity $r$ of the general beam-splitters $\text{BS}_1$ and $\text{BS}_2$ as below:
\begin{eqnarray}
P_\text{eavesdropping} = \frac{1}{12288} \left(1-r\right)^2 \left( 40 - 23r + 7r^2 \right).
\label{eq:P_eav1}
\end{eqnarray}
This probability $P_\text{eavesdropping}$ is to be $\simeq 0.0019$ when $r=0$. $P_\text{eavesdropping}$ becomes $0$ when $r=1$, which intuitively makes sense as Alice cannot encode her message in this case. When $r=\frac{1}{2}$, the probability $P_\text{eavesdropping}$ is as small as $\simeq 0.0006$, which indicates the impracticality of the eavesdropping. Here, one can also consider a more favorable strategy for the eavesdropping, namely a strategy of disturbing only the deterministic paths since the photon traveling along the superposed paths does not carry any message information. Such a strategy is actually possible if Eve can see Bob's initial $\varphi$-setting. Then, the eavesdropping probability $P_\text{eavesdropping}$ is
\begin{eqnarray}
P_\text{eavesdropping} = \frac{1}{8192} \left(1-r\right)^2 \left( 77 - 6r + r^2 \right),
\label{eq:P_eav3}
\end{eqnarray}
which is a bit higher than Eq.~(\ref{eq:P_eav1}), however still vanishingly small; $P_\text{eavesdropping} \simeq 0.0094$ when $r=0$ and $P_\text{eavesdropping} \simeq 0.0023$ when $r=\frac{1}{2}$. Note here that even in such cases, Eve cannot verify whether or not her eavesdropping was successful. 

We then assume that Eve owns a strategy to estimate the polarizations of Bob's initial photon, partially relaxing the complication ({\bf C.1}). Such a clever Eve may try to extract the information of the initial polarizations by looking at the {\em classical} parameters (the angles of the wave-plates) or softwares of Bob's devices $\hat{U}_\text{B}$. In such a case, nearly all existing QSDC (or similar quantum cryptographic) protocols based on the single-photon carriers may become insecure. However, our scheme is not damaged due to the use of the superposition path. Actually, we calculate that even for a more powerful Eve who can completely see Bob's initial polarization, the probability of eavesdropping one valid message-bit is still small. More explicitly, the eavesdropping probability $P_\text{eavesdropping}$ is given by
\begin{eqnarray}
P_\text{eavesdropping} = \frac{1}{48} \left(1-r\right)^2 \left( 12 - 7r + 3r^2 \right),
\label{eq:P_eav2}
\end{eqnarray}
which is higher than those in the above case in Eq.~(\ref{eq:P_eav1}), but still quite small. In fact, the probability is as small as $\simeq 0.0481$ when $r=\frac{1}{2}$. The probability becomes $0.25$ when $r=0$, in which the situation is exactly the same as that of the BB84 protocol. When $r=1$, the probability is equal to $0$. Note that Eve also still cannot verify her success in this case. Therefore, it is impractical to eavesdrop on the full message even with the ability to peep at Bob's initial polarization. We note again that such an advantage cannot be found in a typical (classical) way of just adding the possible pathways (in our case, the arms) to confuse Eve. 

For further analysis, we consider the case of a super-Eve who can attack Bob's whole preparation settings (both the initial polarization and $\varphi$-setting). In this case our protocol cannot be considered secure any more. Eve will measure the initial polarization of the photon traveling `{Bob $\to$ Alice}' only for the deterministic path (i.e., when $\varphi=0$ and $\varphi=\pi$), and will try to extract the bit-message by disturbing the returning photon (in the way of `{Alice $\to$ Bob}') with the measurement bases oriented along Bob's initial polarization axis. Such a super-Eve can reproduce the valid outcomes that would have appeared in Bob's analyzer. Thus, in this case, the superposed paths (the options $\varphi=\frac{\pi}{2}$ and $\varphi=\frac{3}{2}\pi$) are useless. However, Alice and Bob can detect even this super-eavesdropper by subtly-modifying the protocol as follows: Alice adds one more option $\vartheta = \frac{\pi}{8}$ to her setting of the message operation as
\begin{eqnarray}
\hat{U}_\text{A, (R or L)} =
\left\{
\begin{array}{ll}
\hat{\openone} & \text{if Alice sets}~\vartheta=\frac{\pi}{2}, \\
\hat{\sigma}_y  & \text{if Alice sets}~\vartheta=\frac{\pi}{4}, \\
\frac{-i}{\sqrt{2}}\left( \begin{smallmatrix} 1 & 1 \\ -1 & 1 \end{smallmatrix}\right)  & \text{if Alice sets}~\vartheta=\frac{\pi}{8}, \\ 
\end{array}
\right.
\label{eq:option_Um_mod}
\end{eqnarray}
 and Bob's analyzer is initially set to be either an H/V or D/A measurement in each trial, instead of matching to the initial setting $\hat{U}_\text{B}$ (see Fig.~\ref{fig:main_scheme}). Following this modified rule, Alice and Bob now have a chance to detect the aforementioned super-Eve. More specifically, in the cases that
\begin{widetext}
\begin{eqnarray}
&& \underset{\text{Bob's initial polarization}}{\underbrace{\ket{H}~\text{or}~\ket{V}}} \xrightarrow{\text{Bob to Alice}} \underset{\hat{U}_\text{A, (R or L)}~\text{with}~\vartheta=\frac{\pi}{8}}{\underbrace{\ket{A}~\text{or}~\ket{D}}} \xrightarrow{\text{Alice to Bob}} \underset{\text{Bob's analyzer}}{\underbrace{\Big( \text{D/A measurement} \Big)}}, \nonumber \\
&& \text{or}~ \underset{\text{Bob's initial polarization}}{\underbrace{\ket{D}~\text{or}~\ket{A}}} \xrightarrow{\text{Bob to Alice}} \underset{\hat{U}_\text{A, (R or L)}~\text{with}~\vartheta=\frac{\pi}{8}}{\underbrace{\ket{H}~\text{or}~\ket{V}}} \xrightarrow{\text{Alice to Bob}} \underset{\text{Bob's analyzer}}{\underbrace{\Big( \text{H/V measurement} \Big)}},
\label{eq:super-Eve_detect}
\end{eqnarray}
\end{widetext}
it is inferred that Bob and Alice can be aware of an eavesdropping attempt, because in this case Eve's measurement (aligned with the original Bob's polarization axis) in the way of `{Alice $\to$ Bob}' [specifically, in the green arrows in Fig.~\ref{fig:tree}(a)] can alter the polarization and can generate the invalid outcomes of Bob's analyzer. Noting that it is possible for a super-Eve to discriminate the photon path of the message-encoding event, it is easily found that 
\begin{eqnarray}
P_\text{eavesdropping} = P'_\text{message} = \frac{2}{3} \times P_\text{message} = \frac{\left( 1-r \right)^2}{6},
\end{eqnarray}
where $P'_\text{message}$ is the probability of the message-encoding events defined in the above modified protocol. Here, $P_\text{message} = \frac{1}{4} \left(1-r\right)^2$ [see Eq.~(\ref{eq:prob_events})] and the factor $\frac{2}{3}$ is introduced due to the modification of the rule, i.e., adding one more option $\vartheta=\frac{\pi}{8}$. However, if there is an extremely strong eavesdropper who can also attack the random analyzer, then the protocol becomes insecure.

\subsection{A practical strategy of using our protocol.} %-------------------------

We here indicate that the above-described type of super-Eve can verify whether one eavesdropped bit is a valid message-bit or not in each trial, and super-Eve can extract several bits of the secret message until she is detected. Thus, on the basis of the above naive analyses, we provide a practical way of using our QSDC protocol to maximize security advantages. Our basic idea is to identify the existence of Eve as thoroughly as Alice can, prior to communicating the secret message-bits. More explicitly, Alice performs the following {\em beforehand} tests: [{\bf S.1}] First, Alice sets the reflectivity $r$ of her beam-splitters $\text{BS}_1$ and $\text{BS}_2$ to be $1$ without transmitting the photons. Then, all photons are used to detect Eve. However, Alice and Bob cannot detect super-Eve who can look at both the initial polarization and $\varphi$-setting. [{\bf S.2}] Thus, after the completion of the aforementioned stage, Alice retunes $r$ to be $0$ and transmits all photons toward her setting of the message operation $\hat{U}_\text{A, (R or L)}$. Here, Alice chooses only the third $\vartheta=\frac{\pi}{8}$ option in Eq.~(\ref{eq:option_Um_mod}). Then, Alice and Bob can have a chance to detect super-Eve for the cases of Eq.~(\ref{eq:super-Eve_detect}). To make such a strategy even more effective, Alice adopts these two stages [{\bf S.1}] and [{\bf S.2}], by one following the other, such that
\begin{eqnarray}
\underset{T_1}{\underbrace{\text{[\bf S.1]}}} \to \underset{T_2}{\underbrace{\text{[\bf S.2]}}} \to \underset{T_3}{\underbrace{\text{[\bf S.1]}}} \to \cdots \to \underset{T_n}{\underbrace{\text{[\bf S.2]}}},
\label{eq:beforehand_p}
\end{eqnarray}
where $T_j$ ($j=1,2,\ldots,n$) is the time for each stage determined {\em arbitrarily} by Alice. Alice commences the message-bit communication with the two options of $\hat{U}_\text{A, (R or L)}$, as in Eq.~(\ref{eq:option_Um}), and arbitrarily chosen $r \in [0, 1)$, only after verification of the absence of any (super) eavesdropper in the sequential application of [{\bf S.1}] and [{\bf S.2}], as in Eq.~(\ref{eq:beforehand_p}).

%--------------------------------------------------------------------------------------------
\section*{DISCUSSION}
%--------------------------------------------------------------------------------------------

We have presented a general idea for achieving preparation-attack immunity by designing and proposing a novel linear-optical QSDC scheme, called preparation-attack-immune QSDC. In this protocol, a single-photon generated from Bob's side is to move through one of the two pathways (R-arm and L-arm) deterministically, or to simultaneously travel along the superposition path. This photon can be detected in Alice's side, or be reflected (without being assisted by quantum memory) to Bob after taking a bit of Alice's message. In the former case Alice and Bob would share the information of the detection to sense a possible external attack, and in the latter case Bob would decode the bit of Alice's message. We then argued that our protocol is secure in the sense that it would be probabilistically unsuccessful and impractical for any eavesdropper Eve to extract even a piece of message information due to the complications ({\bf C.1})-({\bf C.3}). According to our `naive' analysis, the probability that Eve successfully eavesdrops on one bit of the message is less than $\simeq 0.002$. The remarkable feature was that Alice and Bob can detect the existence of even a more powerful Eve who can attack Bob's initial polarization. In this case where Eve can look at the initial polarization encoding, the eavesdropping probability is $\simeq 0.048$. Analyzing further, if there is a super-Eve who can attack the whole preparation apparatus (both initial polarization and $\varphi$-setting), it is possible to detect her eavesdropping via the subtly-modified rules. Such a new kind of advantage that is the preparation-attack immunity (as opposed to doubling the encoding capacity or patching-up the information leakage \cite{Liu12,Wang15}) was enabled owing to the use of the two-fold photon degree of freedom which maximizes the single-photon quantum superposition nature. In fact, we argued that it is impossible to achieve the preparation-attack immunity just by adding up the effective arms. To our knowledge, this is the first study to the problem of achieving the immunity against the preparation-attack in the preparation-and-measure QSDC scenario. We believe that our scheme can be realized with the current linear-optical stuffs, and further provides a strong motivation to develop the relevant optical technique~\cite{Saitoh16,Dynes16}. Our original idea can be generalized and developed further to improve the security of other single-photon based QKD protocols.

%--------------------------------------------------------------------------------------------

{\em Acknowledgements.} -- We are grateful to Jinhyoung Lee for helpful comments. We also thanks Sang-Min Lee for discussions on experimental realization of the scheme. We acknowledge the financial support of the Basic Science Research Program through the National Research Foundation of Korea (NRF) grant funded by the Ministry of Science, ICT R\&D program of MSIP/IITP (No. R0190-15-2030). JB also acknowledge the financial support of the Basic Science Research Program through the National Research Foundation of Korea (NRF) funded by the Ministry of Science, ICT \& Future Planning (No. 2014R1A2A1A10050117) and the ICT R\&D program of MSIP/IITP (No. 10043464).

\begin{appendix}

\section{Event probabilities}%------------------------

\begin{figure}[h]
\centering
\includegraphics[width=0.4\textwidth]{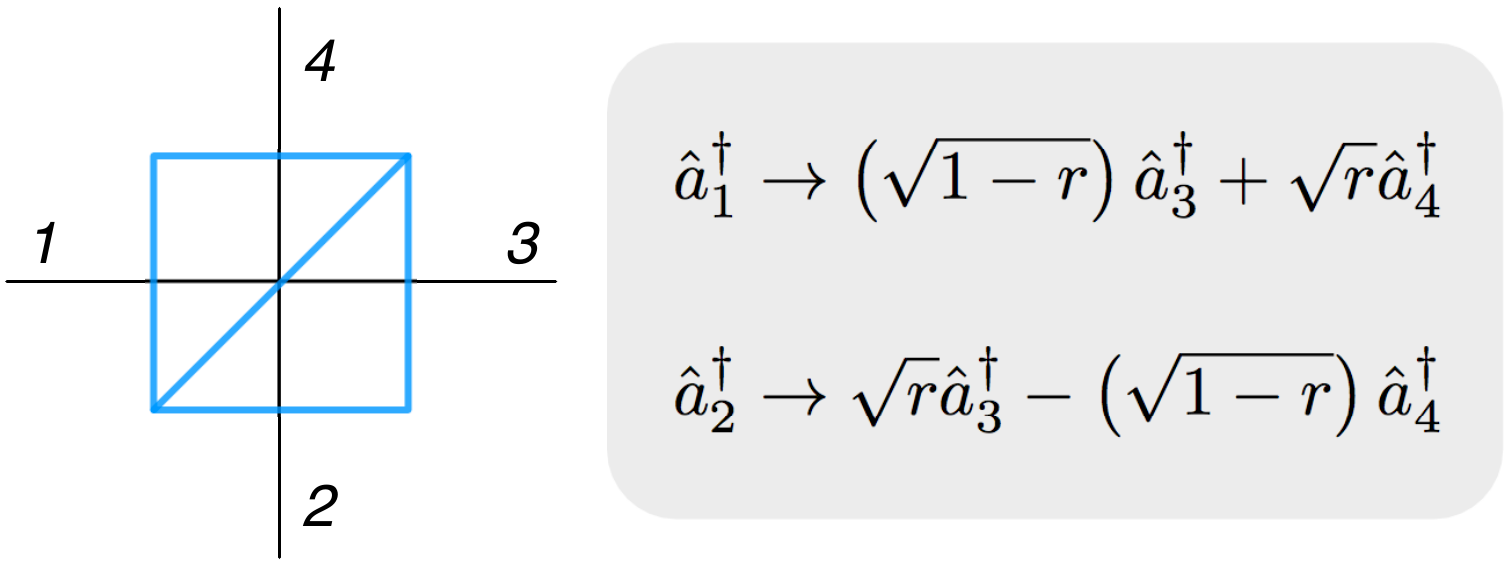}
\caption{\label{fig:BS_convention} The mapping of the creation operators $\hat{a}_j^\dagger$ ($j=1,2,3,4$) in the general beam splitter.}
\end{figure}

Detailed calculations of the event probabilities in Eq.~(6) are presented. First, let us recall the mapping Eq.~(5) of the photon creation operators in each spacial mode:
\begin{eqnarray}
\hat{a}_{\text{T-BS,R}}^\dagger &\to& \left(1-r\right) \times \left(\text{\small - reflected toward Bob \& measured -}\right) \nonumber \\
&& + \sqrt{r \left( 1-r \right)} \hat{a}_{\text{BS}_1,\text{D}_\text{A,1}}^\dagger + \sqrt{\frac{r}{2}} \left( \hat{a}_{\text{BS}_3,\text{D}_\text{A,3}}^\dagger - \hat{a}_{\text{BS}_3,\text{D}_\text{A,4}}^\dagger \right)~\left(\varphi=0\right), \nonumber \\
\frac{\left(\hat{a}_{\text{T-BS,R}}^\dagger + \hat{a}_{\text{T-BS,L}}^\dagger\right)}{\sqrt{2}} &\to& \left(1-r\right) \times \left(\text{\small - reflected toward Bob \& discarded -}\right) \nonumber \\
&& + \sqrt{\frac{r\left(1-r\right)}{2}} \left( \hat{a}_{\text{BS}_1,\text{D}_\text{A,1}}^\dagger + \hat{a}_{\text{BS}_2,\text{D}_\text{A,2}}^\dagger \right) - \sqrt{r} \hat{a}_{\text{BS}_3,\text{D}_\text{A,4}}^\dagger~\left(\varphi=\frac{\pi}{2}\right), \nonumber \\
\frac{\left(\hat{a}_{\text{T-BS,R}}^\dagger - \hat{a}_{\text{T-BS,L}}^\dagger\right)}{\sqrt{2}} &\to& \left(1-r\right) \times \left(\text{\small - reflected toward Bob \& discarded -}\right) \nonumber \\
&& - \sqrt{\frac{r\left(1-r\right)}{2}} \left( \hat{a}_{\text{BS}_1,\text{D}_\text{A,1}}^\dagger - \hat{a}_{\text{BS}_2,\text{D}_\text{A,2}}^\dagger \right) - \sqrt{r} \hat{a}_{\text{BS}_3,\text{D}_\text{A,3}}^\dagger~\left(\varphi=\frac{3}{2}\pi\right), \nonumber \\
\hat{a}_{\text{T-BS,L}}^\dagger &\to& -\left(1-r\right) \times \left(\text{\small - reflected toward Bob \& measured -}\right) \nonumber \\
&& + \sqrt{r \left( 1-r \right)} \hat{a}_{\text{BS}_2,\text{D}_\text{A,2}}^\dagger - \sqrt{\frac{r}{2}} \left( \hat{a}_{\text{BS}_3,\text{D}_\text{A,3}}^\dagger + \hat{a}_{\text{BS}_3,\text{D}_\text{A,4}}^\dagger \right)~\left(\varphi=\pi\right). 
\label{eq:BS_mapping}
\end{eqnarray}
where $\hat{a}_{\text{T-BS,R}}^\dagger$ and $\hat{a}_{\text{T-BS,L}}^\dagger$ create a photon as the output of T-BS traveling along R-arm and L-arm; $\hat{a}_{\text{BS}_j,\text{D}_{\text{A},k}}^\dagger$ denotes the operator to create an output photon from $\text{BS}_j$ ($j=1,2,3$) toward $\text{D}_{\text{A},k}$ ($k=1,2,3,4$). Here, we adopted the beam-splitter convention as in Fig.~\ref{fig:BS_convention}.

\begin{figure}[t]
\centering
\includegraphics[width=0.45\textwidth]{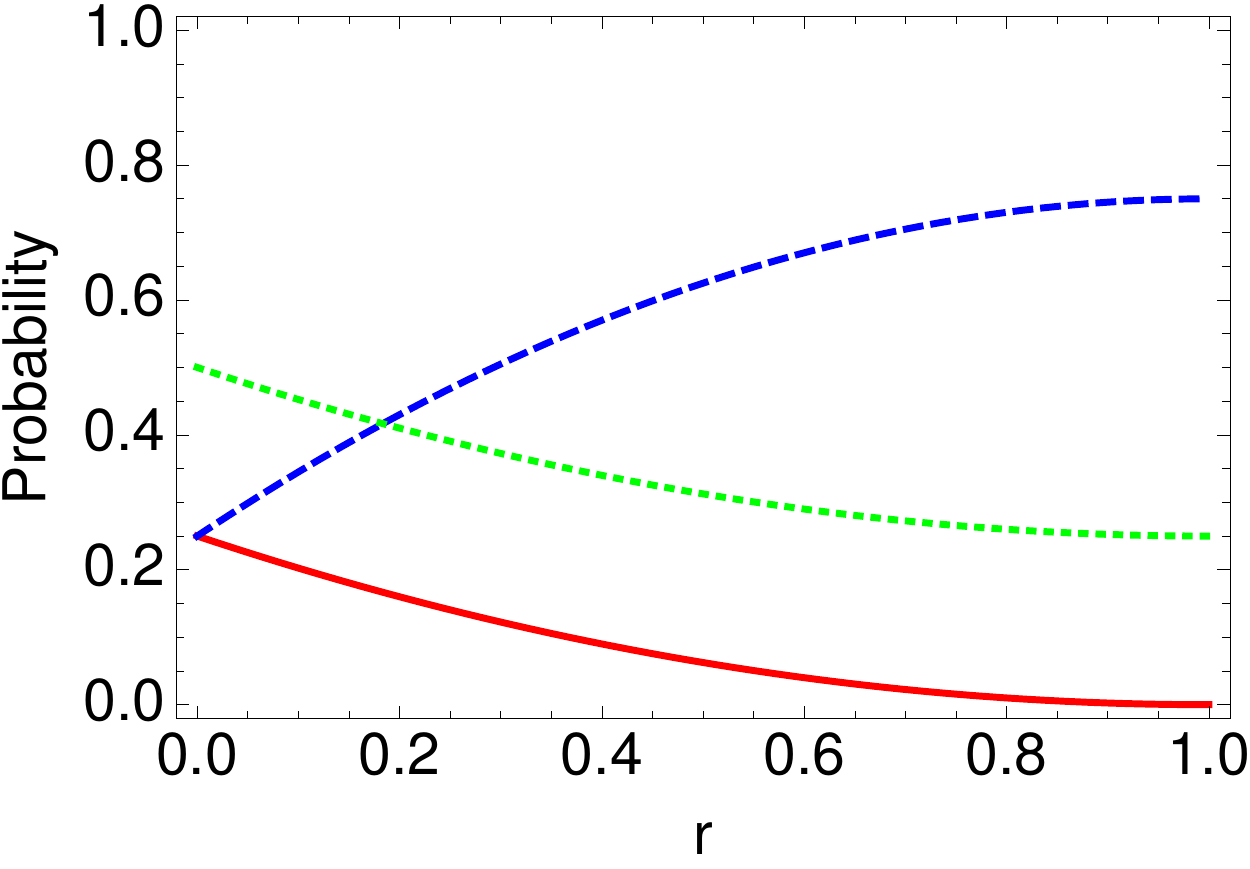}
\caption{\label{grp:prob_events} The graph of the event probabilities $P_\text{message}$ (red solid), $P_\text{Eve-check}$ (blue dashed), and $P_\text{discard}$ (green dotted) are given with respect to $r$.}
\end{figure}

We then describe all possible events and the probabilities of occurrence as below.
\begin{itemize}
\item When $\varphi = 0$ (or $\varphi = \pi$):
\begin{eqnarray}
\left\{
\begin{array}{ccc}
\text{\small Event} & \text{\small Occurring Probability}  & \text{\small Uses}  \\
\text{\small Click Bob's Analyzer} & \frac{(1-r)^2}{2} / \frac{(1-r)^2}{2} & \text{\small Message Decoding / Eve-Check} \\
\text{\small Click $\text{D}_\text{A,1}$ (or $\text{D}_\text{A,2}$)} & r(1-r) & \text{\small Eve-Check} \\
\text{\small Click $\text{D}_\text{A,3}$, $\text{D}_\text{A,4}$} & \frac{r}{2} / \frac{r}{2} & \text{\small Eve-Check / Discarded} 
\end{array}
\right. \nonumber
\end{eqnarray}

\item When $\varphi = \frac{\pi}{2}$ (or $\varphi=\frac{3}{2}\pi$):
\begin{eqnarray}
\left\{
\begin{array}{ccc}
\text{\small Event} & \text{\small Occurring Probability}  & \text{\small Uses}  \\
\text{\small Toward Bob's analyzer} & (1-r)^2 & \text{\small Discarded} \\
\text{\small Click $\text{D}_\text{A,4}$ (or $\text{D}_\text{A,3}$)} & r & \text{\small Eve-Check} \\
\text{\small Click $\text{D}_\text{A,1}$, $\text{D}_\text{A,2}$} & \frac{r(1-r)}{2} / \frac{r(1-r)}{2} & \text{\small Eve-Check / Discarded}
\end{array} 
\right. \nonumber
\end{eqnarray}
\end{itemize}

By observing the events and the occurring probabilities described above, we can calculate the event probabilities as
\begin{eqnarray}
P_\text{message} &=& \frac{1}{4} \times 2 \times \frac{\left(1-r\right)^2}{2} = \frac{\left(1-r\right)^2}{4}, \nonumber \\
P_\text{Eve-check} &=& \frac{1}{4} \times 2 \times \left\{ \left[ r(1-r) + \frac{r}{2} + \frac{(1-r)^2}{2} \right] + \left[ \frac{r(1-r)}{2} + r \right] \right\} = \frac{1 + 4r - 2r^2} {4}, \nonumber \\
P_\text{discard} &=& \frac{1}{4} \times 2 \times \left\{ \frac{r}{2} + \left[ \frac{r(1-r)}{2} + (1-r)^2 \right] + \right\} = \frac{2 - 2r + r^2}{4},
\end{eqnarray}
in which we assumed that the probabilities of choosing $\varphi \in \{ 0, \frac{\pi}{2}, \frac{3\pi}{2}, \pi \}$ are all $\frac{1}{4}$. Note that these event probabilities can be determined by dialing the reflectivity $r$ (or equivalently, the transmittance $1-r$) of Alice's beam splitters $\text{BS}_{1}$ and $\text{BS}_{2}$. This means that the overall time for the completion of the task and the security level can simultaneously be controlled by Alice (see the main manuscript). In Fig.~\ref{grp:prob_events}, we present the event probabilities with respect to $r$.

\section{Eve's successful eavesdropping probabilities}%------------------------

We now present a detailed analysis of the eavesdropping probabilities Eq.~(7), Eq.~(8), and Eq.~(9). To do this, we first calculate the probabilities of Eve not being discovered by considering the following three cases. Eve can be discovered her eavesdropping in the ways of ($i$) {\small Bob $\to$ Alice} for superposed path, ($ii$) {\small Bob $\to$ Alice} for deterministic path, and ($iii$) {\small Alice $\to$ Bob} for deterministic path (we do not need to consider ``{\small Alice $\to$ Bob} for superposed path, because the photon of this path does not carry the message-bit). 

\begin{figure}[t]
\centering
\includegraphics[width=0.45\textwidth]{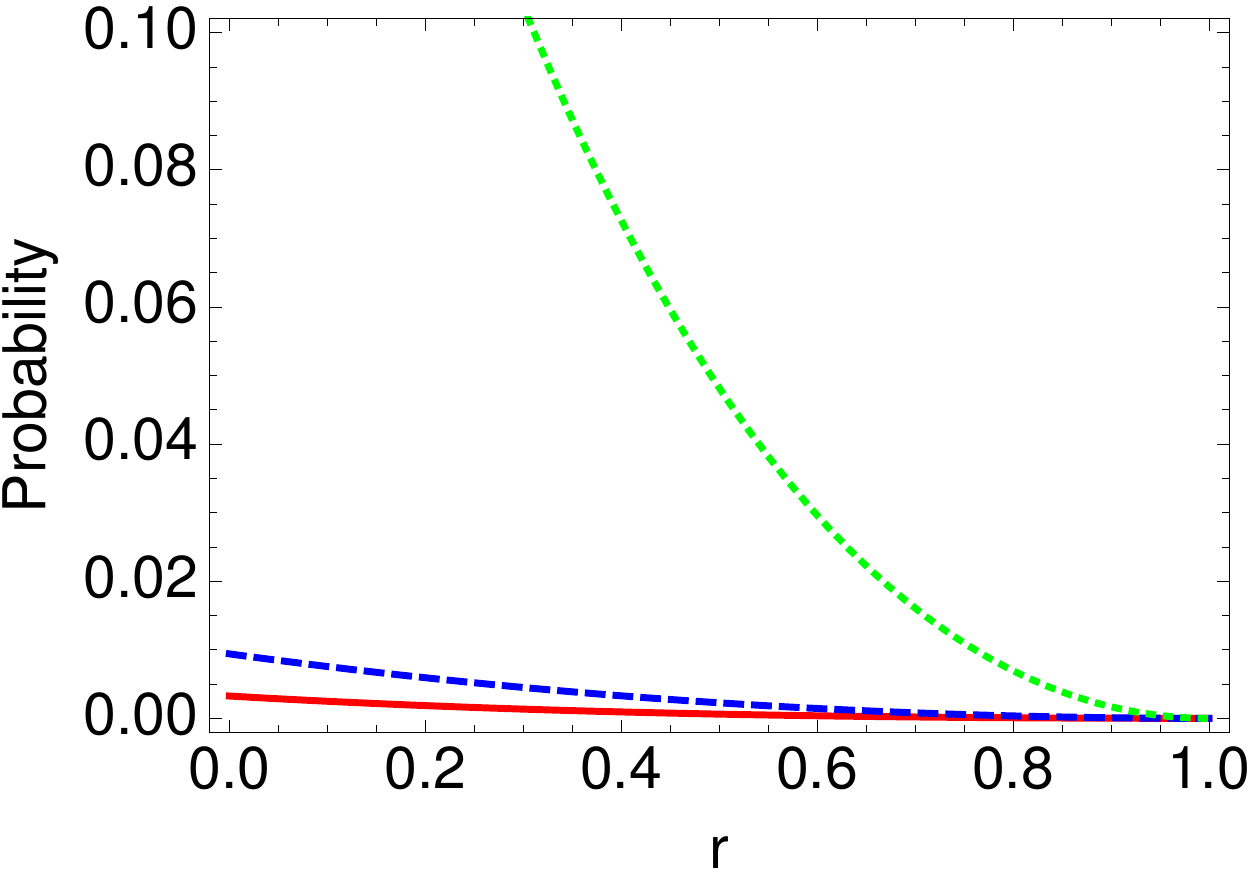}
\caption{\label{grp:prob_eve} The probabilities that Eve can get a bit of Alice's message are given with respect to $r$. The red solid and blue dashed lines denotes the Eve's successful eavesdropping probabilities for the cases that Eve cannot get any information and Eve can look at only $\varphi$-setting, respectively. The green dotted line is the probability given for Eve who can directly attack Bob's polarization.}
\end{figure}  

{\em Scenario $1$.} -- First, let us consider the case that Eve does not have any information regarding Bob's preparation apparatus. In such a case, we can find that 
\begin{eqnarray}
\text{\small The probability of Eve not being discovered} = 
\left\{
\begin{array}{ll}
\frac{1}{384} \left( 41 - 28r + 11r^2 \right) & ~ \text{for ($i$)}, \\
\frac{1}{32} \left( 5r - 3r^2 \right)               & ~ \text{for ($ii$)}, \\
\frac{13}{128} \left( 1-r \right)^2                & ~ \text{for ($iii$)},
\end{array}
\right.
\end{eqnarray}
where all possible events are considered (refer to Fig.~2 in the main manuscript). Thus, the overall probability that Eve is not detected is given, such that $(i) + (ii) + (iii) = \frac{1}{192} (40 - 23r + 7r^2)$. However, noting that Eve's successful eavesdropping is valid only for the message-encoding events, we can find the eavesdropping probability $P_\text{eavesdropping}$ as
\begin{eqnarray}
P_\text{eavesdropping} &=& \frac{1}{16} \times P_\text{message} \times \frac{1}{192} (40 - 23r + 7r^2) \nonumber \\
     &=& \frac{1}{12288} (1-r)^2 (40 - 23r + 7r^2),
\label{eq:Peve_c1}
\end{eqnarray}
where the factor $\frac{1}{16}$ is introduced because Eve cannot verify both the initial polarization and $\varphi$-setting. 

{\em Scenario $2$.} -- Then, we consider a more powerful Eve who can discriminate the deterministic superposed paths by looking at Bob's $\varphi$-setting. Such an Eve can follow a bit more improved strategy, namely a strategy of disturbing only the photons that travel along the deterministic paths. In this case, Alice and Bob can detect such an Eve with the probabilities as
\begin{eqnarray}
\text{\small The probability of Eve not being discovered} = 
\left\{
\begin{array}{ll}
\frac{1}{2}                             & ~ \text{for ($i$)}, \\
\frac{1}{32} r\left(5-3r\right)  & ~ \text{for ($ii$)}, \\
\frac{13}{128} \left( 1-r \right)^2  & ~ \text{for ($iii$)},
\end{array}
\right.
\end{eqnarray}
and we can evaluate
\begin{eqnarray}
P_\text{eavesdropping} = \frac{1}{16} \times P_\text{message} \times \Big[ (i) + (ii) + (iii) \Big] = \frac{1}{8192} (1-r)^2 (77 - 6r + r^2).
\label{eq:Peve_c2}
\end{eqnarray}
Here we note that $P_\text{eavesdropping}$ is larger than Eq.~(\ref{eq:Peve_c1}), but still vanishingly small. 

{\em Scenario $3$.} -- Lastly, we can also consider another type of powerful Eve who can look at Bob's initial polarization encoding, which is a crucial information of the secret message. In this case, Eve can align her own measurement polarization setting of $\text{D}_\text{Eve,R}$ and $\text{D}_\text{Eve,L}$ according to Bob's original encoding. Such an Eve can also make the valid outcomes appear in Bob's analyzer. Thus we can analyze that
\begin{eqnarray}
\text{\small The probability of Eve not being discovered} = 
\left\{
\begin{array}{ll}
\frac{1}{2}                             & ~ \text{for ($i$)}, \\
\frac{1}{12} r\left(5-3r\right)  & ~ \text{for ($ii$)}, \\
\frac{1}{2} \left( 1-r \right)^2  & ~ \text{for ($iii$)},
\end{array}
\right.
\end{eqnarray}
and
\begin{eqnarray}
P_\text{eavesdropping} = P_\text{message} \times \Big[ (i) + (ii) + (iii) \Big] = \frac{1}{48} (1-r)^2 (12 - 7r + 3r^2),
\end{eqnarray}
which is much larger than Eq.~(\ref{eq:Peve_c1}) and Eq.~(\ref{eq:Peve_c2}). Here we indicate that when $r=0$, Eve has $P_\text{eavesdropping} = 0.25$. This intuitively makes sense, as in this case the situation is exactly equal to that of BB84 for Eve. Note further that $P_\text{eavesdropping}$ becomes small when $r \to 1$. Of course, in this case, the overall time to complete the task will increase (see the main manuscript). In Fig.~\ref{grp:prob_eve}, we depict the above described probabilities with respect to $r$. 

\end{appendix}

\bibliographystyle{naturemag}

\end{document}